# Unconventional temperature enhanced magnetism in iron telluride

Igor A. Zaliznyak[1], Zhijun Xu[1], John M. Tranquada[1], Genda Gu[1], Alexei M. Tsvelik[1], Matthew B. Stone[2]

[1] *CMP&MS Department, Brookhaven National Laboratory, Upton, New York 11973, USA*

[2] *Oak Ridge National Laboratory, 1, Bethel Valley Road, Oak Ridge, Tennessee 37831, USA*

**Discoveries of copper and iron-based high-temperature superconductors (HTSC)[1-2] have challenged our views of superconductivity and magnetism. Contrary to the pre-existing view that magnetism, which typically involves localized electrons, and superconductivity, which requires freely-propagating itinerant electrons, are mutually exclusive, antiferromagnetic phases were found in all HTSC parent materials[3,4]. Moreover, highly energetic magnetic fluctuations, discovered in HTSC by inelastic neutron scattering (INS) [5,6], are now widely believed to be vital for the superconductivity [7-10]. In two competing scenarios, they either originate from local atomic spins[11], or are a property of cooperative spin-density-wave (SDW) behavior of conduction electrons [12,13]. Both assume clear partition into localized electrons, giving rise to local spins, and itinerant ones, occupying well-defined, rigid conduction bands. Here, by performing an INS study of spin dynamics in iron telluride, a parent material of one of the iron-based HTSC families, we have discovered that this very assumption fails, and that conduction and localized electrons are fundamentally entangled. In the temperature range relevant for the superconductivity we observe a remarkable redistribution of magnetism between the two groups of electrons. The effective spin per Fe at T $\approx$ 10 K, in the**

**antiferromagnetic phase, corresponds to S ≈ 1, consistent with the recent analyses that emphasize importance of Hund's intra-atomic exchange[15-16]. However, it grows to S ≈ 3/2 in the *disordered* phase, a result that profoundly challenges the picture of rigid bands, broadly accepted for HTSC.**

Iron telluride is the parent material of the simplest family of iron-based superconductors, and can be made superconducting by partial, or full isoelectronic substitution of Te by Se[17,18]. Although the highest critical temperature for $FeTe_{1-x}Se_x$ is only $T_c \approx 14.5$ K, it increases to above 30 K under pressure[19]. The crystal structure of iron telluride consists of a continuous stacking of FeTe layers, which represent the basic structural motif of all iron-based superconductors, where a square-lattice layer of iron atoms is sandwiched between two twice-sparser layers of bonding chalcogen/pnictogen atoms. The Te atoms, which tetrahedrally coordinate the Fe sites, occupy alternate checkerboard positions above and below the Fe layer, so that the resulting unit cell contains two formula units. In this quasi-two-dimensional structure, FeTe layers are held together only by weak Van der Waals forces. Crystallographic stability is improved if some amount of extra Fe atoms is incorporated between the layers, which frustrates magnetic correlations in the $Fe_{1+y}Te$ series ($0.02 < y < 0.11$), as we will discuss later.

Band structure calculations predict Fe(Te,Se) to be a metal with several bands crossing the Fermi energy[20-22]. This qualitatively agrees with scanning tunnelling spectroscopy[23] and angle-resolved photoemission measurements of FeTe[24,25], which both find small electron and hole pockets near the corner, $\mathbf{Q} = (\pi,\pi)$ and center, $\mathbf{Q} = 0$, of the two-dimensional (2D) Brillouin zone (BZ), respectively (we use notations corresponding to the actual crystallographic unit cell containing 2 Fe atoms). While these findings unambiguously reveal the existence of itinerant electrons, bulk resistivity measurements find either a non-metallic, or bad-metal behaviour, at best. At the same time, Curie-Weiss behaviour of magnetic susceptibility suggests significant local magnetic moments, $\mu_{eff} \approx 4\mu_B$ ($\mu_B$ = Bohr's magneton), and a rather small Curie-Weiss temperature, $\Theta_{C-W} \leq 190$ K[28]. Thus, we have materials where local moments and itinerant conduction electrons coexist, and the connections between them are revealed by our experiments.

Figure 1 presents an overview of the low-energy magnetic excitations which we have measured in a large single crystal of $Fe_{1.1}Te$ at T = 10 K. The left column shows neutron intensity as a function of the 2D wave vector $\mathbf{Q} = 2\pi(h,k)$ in the *ab*-plane for elastic, E = 0, (a), and inelastic, E = 7.5 meV, (b), and 20 meV, (c), scattering. Remarkably, the scattering takes the form of broad, diffuse peaks centered near (but not exactly at) $(\pm\pi,0)$ and $(0,\pm\pi)$ positions, for all energies covered in this measurement (E $\leq$ 26 meV). Magnetic dynamics of this type is often explained by invoking a system of itinerant electrons, where wave vectors of magnetic excitations are determined by nesting properties of the Fermi surface(s)[12-13,20]. Such explanation clearly fails for FeTe compounds, since there is no Fermi surface nesting near $(\pi,0)$ – nesting occurs at $(\pi,\pi)$ and $(2\pi,0)$[23]. In addition, as we show later, the large observed magnetic intensity would require the entire weight of two fully spin-polarized itinerant electronic bands.

The magnetic excitations in Fig. 1 imply robust short-range correlations, whose well-defined real space structure persists over a broad range of time scales. In a system of local spins this might be a signature of an emergent cooperative spin texture which governs low-energy excited states, such as the hexagonal loops induced by spin frustration in $ZnCr_2O_4$ [30]. We have investigated this conjecture for a number of cluster models and find that magnetic scattering in $Fe_{1.1}Te$ can indeed be very accurately described by a local spin model of this kind. The right column of Fig. 1 shows fits of our data to a model in which plaquettes of four ferromagnetically co-aligned nearest-neighbour Fe spins emerge as a new collective degree of freedom, with short-range antiferromagnetic correlations between neighboring plaquettes. The absence of magnetic scattering along the sides of the square with vertices at (h,k) = $(\pm1,0)$, $(0,\pm1)$ is a fingerprint of the plaquette form factor, $S_p(\mathbf{Q}) \sim |\cos(\pi(h+k)/2)\cos(\pi(h-k)/2)|^2$. With only two parameters, the intensity and the correlation length $\xi$, this fit is nearly as good as the fit to the phenomenological pattern of factorized Lorentzian peaks shown in Fig. 1 (d)-(f), which was used for quantifying intensity and position of magnetic scattering in Figures 2 and 3.

Figure 2 shows the energy dependence of the magnetic intensity at $h \approx 0$, corresponding to the vertical slice at the center of Fig. 1(a)-(c). It reveals striking resonant behaviour, with a pronounced maximum at $E \approx 7$ meV and a weak, acoustic-like mode dispersing from the resonance down to $Q = 0$. While the origin of the resonance is unclear, it appears at the same energy as the spin resonance emerging at $(\pi,\pi)$ in superconducting $FeTe_{1-x}Se_x$ samples[6]. Our supplementary measurements have revealed that magnetic scattering pattern in $Fe_{1.1}Te$ begins changing noticeably above ~ 30 meV, and at high energies has similar shape to that in $FeTe_{1-x}Se_x$ (see Ref. 6), adding to the tantalizing similarity. We find that magnetic fluctuations in $Fe_{1.1}Te$ extend to $\approx$ 190 meV, an order of magnitude larger than $k_B\Theta_{C\text{-}W}$. This is a clear signature of competing interactions. Indeed, the former scale pertains to the energy associated with a flip of a single spin thus giving a typical value of the exchange coupling, while $\Theta_{C\text{-}W}$ is determined by the sum of all exchange couplings, and its relative smallness in $Fe_{1+y}Te$ indicates an almost complete cancellation of the dominant next-nearest neighbour Fe-Te-Fe antiferromagnetic superexchange and the ferromagnetic coupling between the nearest neighbours, which is probably mediated by itinerant electrons.

As far as the low energy magnetic fluctuations are concerned, it is tempting to pursue further the analogy with emergent excitations of frustrated spins in $ZnCr_2O_4$, where somewhat similar resonance behaviour was also observed. There, it arises upon cooling as a result of interaction with the lattice, and is accompanied by a small lattice distortion and weak long-range antiferromagnetic order[30]. In fact, $Fe_{1+y}Te$ materials also exhibit magnetic order, which is coincident with small monoclinic distortion of the crystal lattice[26,27]. At high temperatures both systems have similar effective Curie-Weiss fluctuating moments of ~4$\mu_B$, and in both cases only a fraction of the moment participates in the long-range antiferromagnetic order.

Magnetic order and lattice distortion in our $Fe_{1.1}Te$ sample are weaker than in samples with lower y ($\leq 0.06$)[26]. The ordering occurs at $T_N \approx 58$ K, about 10 K lower than at lower y, and without the sharp discontinuity in magnetic susceptibility (see supporting material), which suggests that the Fe interstitials contribute to magnetic frustration. Analysis of magnetic Bragg intensities, which are visible in the 10 K data of Fig. 1(a) near ($\pm 1$, $\pm 0.5$), yields a long range ordered moment $\approx 1\mu_B$, about twice smaller than for y = 0.05[26,27]. However, there is also a significant amount of elastic or nearly elastic 2D magnetic diffuse scattering near (0,$\pm 0.5$), ($\pm 0.5$,0) in our sample, which results from frozen, or very slowly fluctuating, short-range correlations between the plaquettes. The total spectral weight of magnetic Bragg and diffuse scattering adds to $\approx 2.3\mu_B$, which is similar to the ordered moment in samples with lower y.

Recall, that for a system of spins S, magnetic neutron scattering intensity is determined by the product of magnetic form factor, which accounts for electronic magnetization density associated with each spin, and the dynamical correlation function S(**Q**, E), which describes cooperative motions of spins. The total spectral weight of S(**Q**, E) obeys the sum rule, $\int S(Q,E)d^3QdE = S(S+1) = [\mu_{eff}/(g\mu_B)]^2$, where we omit polarization indices and imply trace over spin polarizations; g is the Lande factor. This defines the fluctuating instantaneous effective moment, $\mu_{eff}$, whereas the ordered static moment is $g\mu_B\langle S\rangle$, where $\langle S\rangle$ is the ground-state value of spin-S operator, with $g\mu_B\langle S\rangle < g\mu_B S < \mu_{eff}$. Figure 3a shows that magnetic Bragg peaks account for only $\approx 28\%$ of the total intensity, while another $\approx 28\%$ is in the inelastic spectrum. By energy-integrating all contributions, we obtain $\mu_{eff} = 2.7(7)$ $\mu_B$ at 10 K, which is very close to $\approx 2.8$ $\mu_B$ expected for S = 1 and g = 2.

Thus, only about a half of the total magnetic intensity expected for $\mu_{eff} \approx 4\mu_B$ obtained from the uniform CW susceptibility is accounted in our T = 10 K data. This already suggests that we do not deal with just a system of local spins. The temperature dependence provides further striking evidence. Indeed, for a system of spins S, the sum rule requires that the integral of S($\mathbf{Q}$, E) remains constant at all temperatures. The magnetic scattering in insulating $ZnCr_2O_4$ is consistent with local moments that do not change with temperature. In $Fe_{1.1}Te$ the behaviour is dramatically different, as shown in Figs. 2 and 3. We find that total magnetic INS intensity significantly increases upon heating, as summarized in Fig. 4(c). The total magnetic spectral weight at 300 K yields $\mu_{eff} \approx 3.6$ $\mu_B$, close to the value of $\approx 3.9$ $\mu_B$ corresponding to S = 3/2 and in good agreement with the susceptibility data. Thus the overall picture is of a temperature-induced change from local spins S = 1 at 10 K to S = 3/2 at 300 K. This can only occur as a result of an effective change by 1 of the number of localized electrons, with a corresponding change in the number of itinerant electrons.

Having made this surprising discovery, we performed a more detailed survey of the temperature dependence of S($\mathbf{Q}$, E). Figure 4(b) shows the resulting wave-vector-integrated correlation function S(E) on a logarithmic scale, which emphasizes the changing balance between quasi-elastic and inelastic fluctuations. As illustrated in Fig. 4(a), the resonance character of the inelastic spectrum in $Fe_{1.1}Te$ is clearly retained even at 300 K, in contrast to the behaviour of the frustrated local spins in insulating $ZnCr_2O_4$. It is this resonance mode which is the main beneficiary of the additional temperature-induced spectral weight in S(E). A possible origin of such an incoherent resonant mode is screening of the local moments by the conduction electrons, as in the Kondo effect.

Finally, as one can see in Fig. 4(c), the sum of the temperature-dependent magnetic Bragg intensity and the quasi-elastic 2D diffuse scattering is nearly T-independent right through $T_N$ and beyond. This suggests that both long-range antiferromagnetism and the monoclinic distortion are likely to be only modest perturbations to the emergent spin dynamics, which is governed by much stronger interactions of local spins and itinerant electrons.

While the relevance of magnetic correlations to HTSC is widely acknowledged[20], the nature of the magnetism has been controversial. Here, for a system bearing an immediate relation to HTSC, we have presented direct experimental evidence for emergent local spin magnetism non-trivially coupled to the itinerant electrons. While the nature of the local spin clusters which govern low-energy magnetic fluctuations in $Fe_{1.1}Te$ is not favourable for the HTSC, perhaps a slight change of the electronic structure in $FeTe_{1-x}Se_x$ modifies the emergent modes and their interaction with itinerant electrons in a manner conductive for the superconductivity. To our knowledge, the temperature-induced enhancement of local magnetic moments that we have found in iron telluride has never before been observed in a magnetic iron group material. This effect presents a new challenge, as it has not been anticipated by any theoretical work, and is not expected in a description based on rigid bands for the conduction electrons.

**Methods**

The $Fe_{1.1}Te$ crystal used in our measurements had a mass of 18.45 g and a mosaic of 2.2° full width at half maximum (FWHM). It was mounted on an aluminium holder attached to the cold head of closed-cycle refrigerator on the ARCS spectrometer at the Spallation Neutron Source at Oak Ridge National Laboratory, USA. The crystal's *c*-axis was aligned parallel to the incident beam, with the *a*-axis at about 24° to the horizontal.

ARCS is a direct geometry, time-of-flight neutron spectrometer, with, a monochromatic pulse of neutrons incident on the sample with energy $E_i$. For each energy transfer E ($< E_i$), a slice of the sample's $\mathbf{Q} = (h, k, l)$ phase space is probed. We analyzed projections of such slices on the (*h,k*)-plane. Each (*h, k*) point corresponds to a particular value of $l = l(h, k, E, E_i)$, since $\mathbf{Q}$ and E are coupled via energy-momentum conservation. Choosing $E_i = 40$ meV we can observe magnetic Bragg peaks arising from 3D magnetic ordering at wave vector $\approx (0.5,0,0.5)$ in the second Brillouin zone, Fig. 1(a), because $l(0.5, 1, 0, E_i = 40) \approx 0.5$. Previous studies of similar compositions have established that diffuse quasi-elastic and inelastic scattering is of a 2D character and has weak or no *l*-dependence, which can be neglected in our analysis.

$\mathbf{Q}$-independent background (BG) arising from incoherent elastic scattering in the sample (same for all T) was subtracted from all data. It was estimated at several positions where magnetic scattering is nearly absent, and had the form of the resolution-limited (FWHM = 2 meV) Gaussian peak in energy. Figures 1 and 2 show the BG subtracted data. Data were normalized through an analysis of the inelastic scattering intensity from acoustic phonon modes near the structural Bragg reflection $\tau = (1, 1, 1)$.

**Acknowledgements** We acknowledge discussions with G. Xu, C. Petrovic, C. Homes, and M. Khodas. Work at BNL was supported by the U.S. Department of Energy.

**Competing Interests statement** The authors declare they have no competing financial interests.

**Correspondence** and requests for materials should be addressed to I. A. Z. (zaliznyak@bnl.gov).

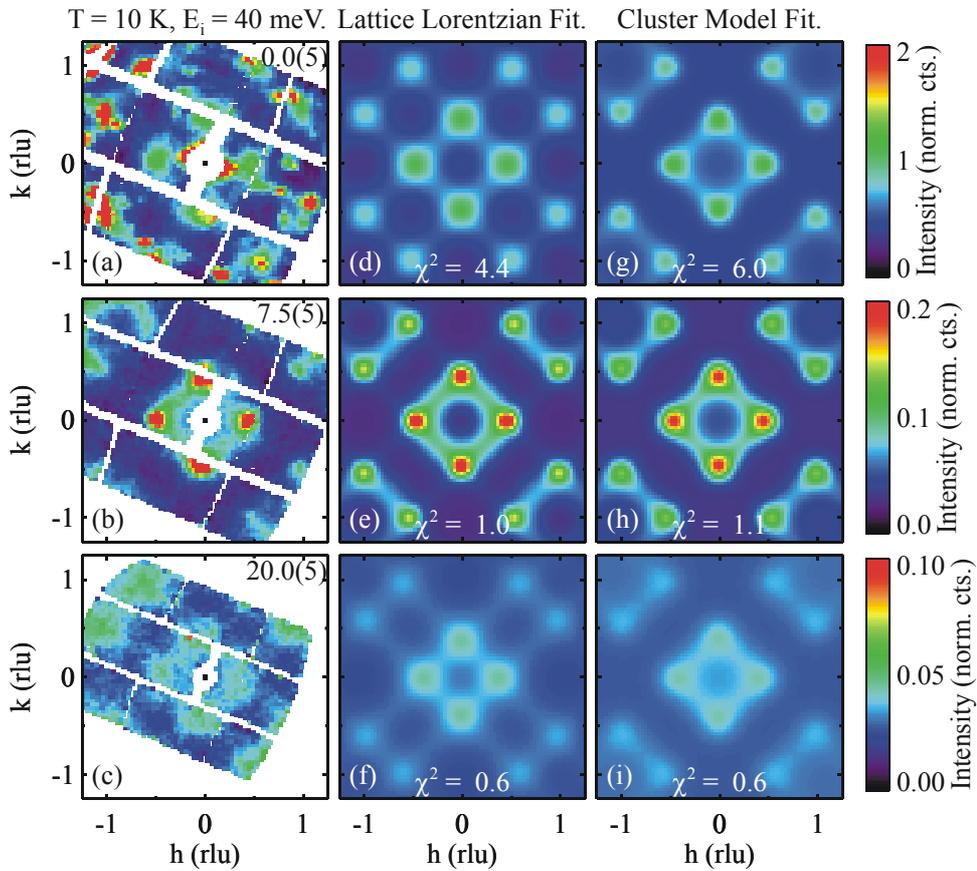

**Figure 1 – Magnetic scattering in Fe$_{1.1}$Te at T = 10 K.** (**a**)-(**c**) Measured intensity for energy transfers (0 ± 0.5) meV, (7.5 ± 0.5) meV and (20 ± 0.5) meV, respectively. (**d**)-(**f**) Fits with a model cross section described by four lattice Lorentzian (LL) peaks at (0,±ς) and (±ς,0), where ς < 0.5. Panel (**e**) also includes a Gaussian ring of scattering centered around (0,0) to account for the dispersive acoustic mode clearly visible in Fig. 2(a). (**g**)-(**i**) Two-parameter fits to the checkerboard cluster model, as described in the text. All fits account for the magnetic form factor of Fe$^{2+}$. Intensity is shown in counts normalized to a unit phase space volume.

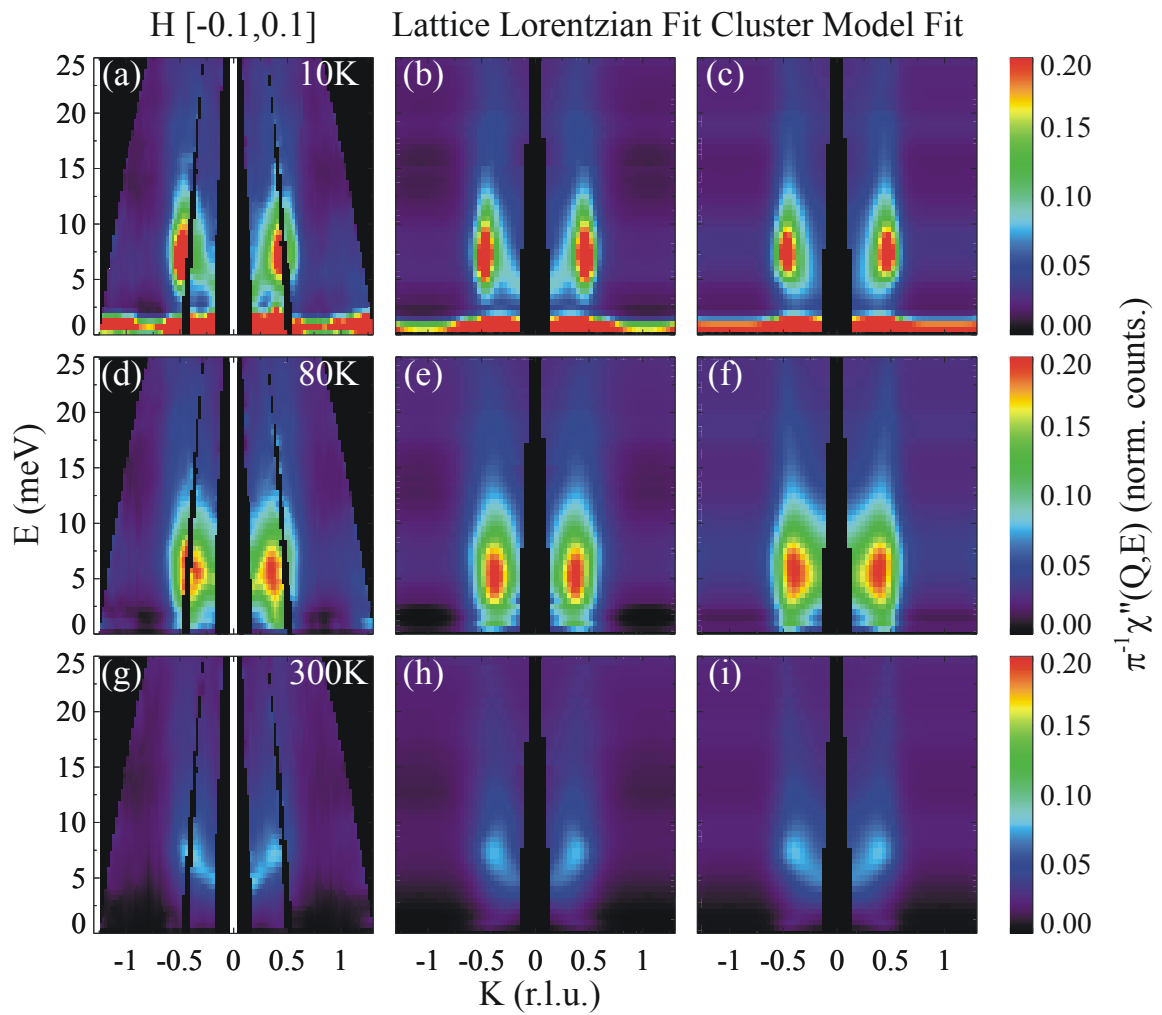

**Figure 2 – Energy dependence of the imaginary part of the dynamical magnetic susceptibility, χ″(Q,E).** (**a**), (**d**), (**g**) Data at T = 10 K, 80 K and 300 K, respectively, plotted as a function of wave vector (0,K). Signal extracted from the intensity using χ″(Q,E) = π(1-e^{-E/(k_BT)})S(Q,E), and shown in normalized counts. The fitted results in (**b**), (**e**), (**h**) and in (**c**), (**f**), (**i**) correspond to the same models as in Fig. 1.

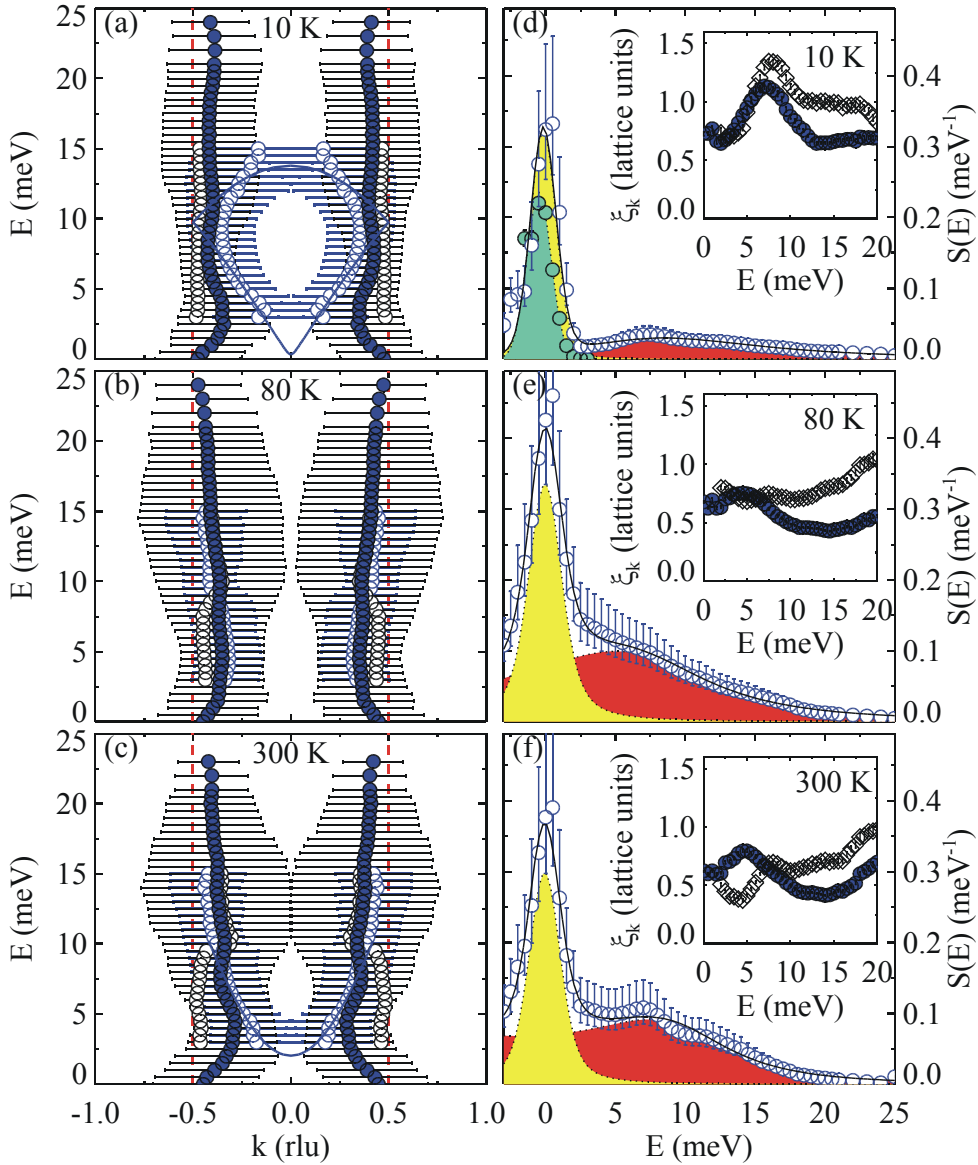

**Figure 3 – Position and intensity of magnetic excitations in Fe$_{1.1}$Te.** Filled symbols in (**a**), (**c**), (**e**) show peak positions at T = 10 K, 80 K and 300 K, respectively, obtained by fitting constant-E slices to a single-component LL cross-section. Horizontal bars show the full width at half maximum (FWHM) of LL peaks. Open symbols are positions of the LL (black) and the ring (blue with error bars) components in the two-component model, as in Fig. 2(b),(e),(h). Solid line in (**a**) shows a fit of the ring mode dispersion to E(q) ~ sin(πk/2), folded into a small (magnetic) Brillouin zone, [-0.5,0.5]. (**b**), (**d**), (**f**) Integral intensity of magnetic scattering as a function of energy. Error bars include the uncertainty of absolute normalization. Solid lines are fits used to interpolate the data for integration. They consist of a quasi-elastic (QE) central peak (shaded yellow) and a damped harmonic oscillator (DHO, red). Green-shaded peak in (**b**) indicates magnetic Bragg intensity. Insets show the correlation length in lattice units for the single-component LL (closed circles) and the cluster (open rhombi) models.

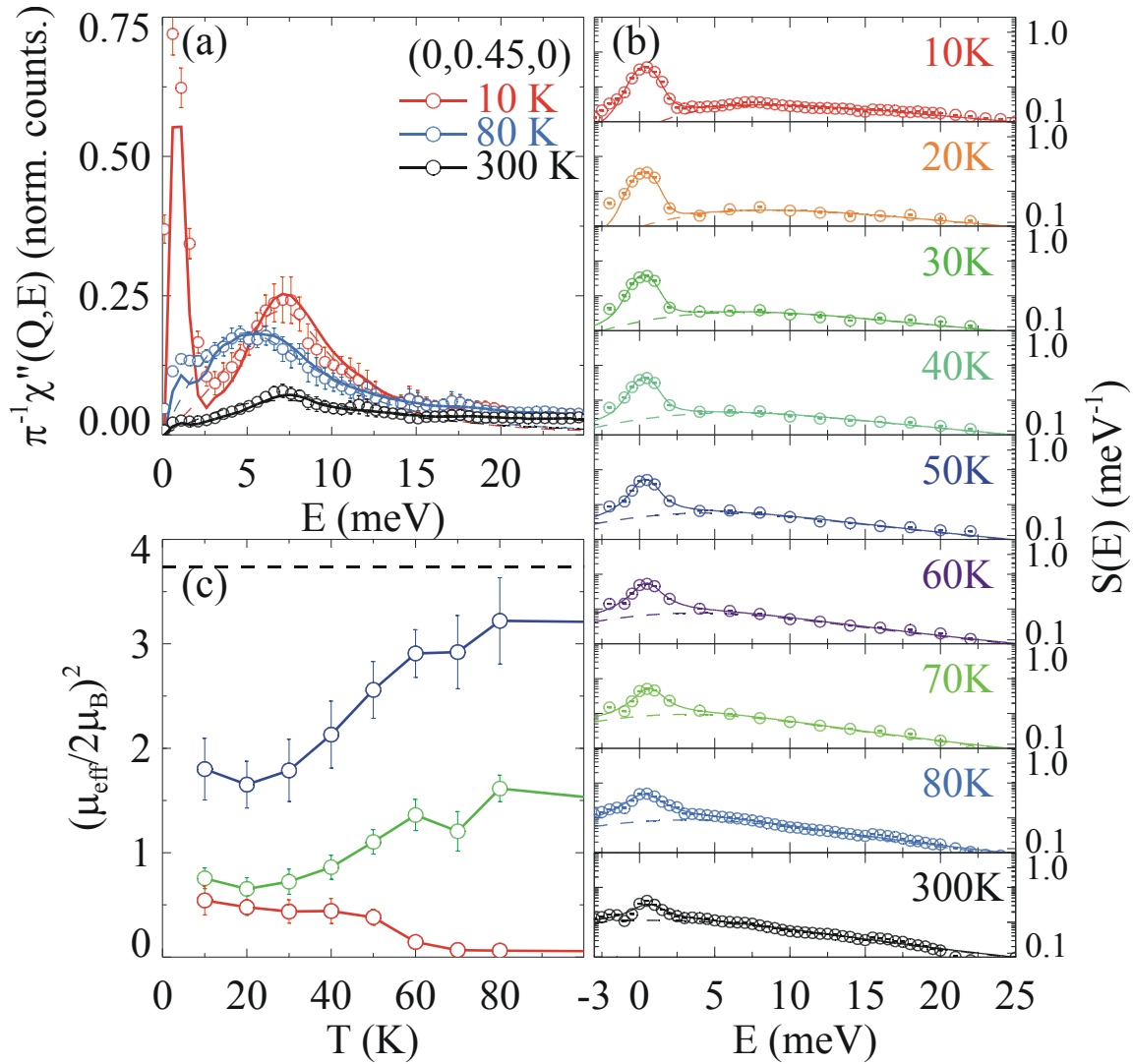

**Figure 4 – Temperature dependence of scattered intensity and $\mu_{eff}^2$.**

(**a**) $\chi''(\mathbf{Q},E)$ as a function of energy for $\mathbf{Q}$ = (0,0.45) at 10 K, 80 K and 300 K. Lines correspond to fits shown in Fig. 2(b),(e),(h). (**b**) Temperature dependence of inelastic and quasielastic magnetic contributions to S(E) (excluding Bragg scattering). Solid lines correspond to the fits of Fig. 3(d)-(f); dashed lines show inelastic contribution modelled by DHO. (**c**) Square of $\mu_{eff}/2\mu_B$ obtained from integrating S(E), shown as a function of temperature. Upper (blue) symbols show total response; bottom (red) symbols indicate the contribution from Bragg peaks; green symbols show contribution from the quasi-elastic scattering.

Horizontal dashed line shows the result obtained from the Curie-Weiss fit of static magnetic susceptibility.